\documentstyle[12pt]{article}

\begin{document}

%%%%%%%%%%%%%%%%%%%%%%%%%%%%%%%%%%%%%%%%%%%%%%%%%%%%%%%%%%%%%%%%%%%
% sugawara's macro
%%%%%%%%%%%%%%%%%%%%%%%%%%%%%%%%%%%%%%%%%%%%%%%%%%%%%%%%%%%%%%%%%%%
\newcommand{\Om}{\Omega}
\newcommand{\df}{\stackrel{\rm def}{=}}
\newcommand{\co}{{\scriptstyle \circ}}
\newcommand{\de}{\delta}
\newcommand{\lb}{\lbrack}
\newcommand{\rb}{\rbrack}
\newcommand{\rn}[1]{\romannumeral #1}
\newcommand{\msc}[1]{\mbox{\scriptsize #1}}
\newcommand{\dsp}{\displaystyle}
\newcommand{\scs}[1]{{\scriptstyle #1}}

\newcommand{\ket}[1]{| #1 \rangle}
\newcommand{\bra}[1]{| #1 \langle}
\newcommand{\vac}{| \mbox{vac} \rangle }

\newcommand{\e}{\mbox{{\bf e}}}
\newcommand{\va}{\mbox{{\bf a}}}
\newcommand{\bc}{\mbox{{\bf C}}}
\newcommand{\br}{\mbox{{\bf R}}}
\newcommand{\bz}{\mbox{{\bf Z}}}
\newcommand{\bq}{\mbox{{\bf Q}}}
\newcommand{\bn}{\mbox{{\bf N}}}
\newcommand {\eqn}[1]{(\ref{#1})}

\newcommand{\cp}{\mbox{{\bf P}}^1}
\newcommand{\n}{\mbox{{\bf n}}}
\newcommand{\sbz}{\msc{{\bf Z}}}
\newcommand{\sn}{\msc{{\bf n}}}

\newcommand{\be}{\begin{equation}}\newcommand{\ee}{\end{equation}}
\newcommand{\bea}{\begin{eqnarray}} \newcommand{\eea}{\end{eqnarray}}
\newcommand{\ba}[1]{\begin{array}{#1}} \newcommand{\ea}{\end{array}}

\newcommand{\cleqn}{\setcounter{equation}{0}}
\makeatletter
\@addtoreset{equation}{section}
\def\theequation{\thesection.\arabic{equation}}
\makeatother

\def\np{Nucl. Phys. {\bf B}}
\def\pl{Phys. Lett. {\bf B}}
\def\mpl{Mod. Phys. {\bf A}}
\def\ijmp{Int. J. Mod. Phys. {\bf A}}
\def\cmp{Comm. Math. Phys.}
\def\prd{Phys. Rev. {\bf D}}

\def\vu{\vec u}
\def\vv{\vec v}
\def\vt{\vec t}
\def\vn{\vec n}
\def\ve{\vec e}
\newcommand{\matriz}[4]{\left(
\begin{array}{cc}#1&#2\\#3&#4\end{array}\right)}

%%%%%%%%%%%%%%%%%%%%%%%%%%%%%%%%%%%%%%%%%%%%%%%%%%%%%%%%%%%%%%%%%%%
\begin{flushright}
La Plata Th-99/01\\April, 1999
\end{flushright}

\bigskip

\begin{center}

{\Large\bf Non standard parametrizations and adjoint invariants\\
of classical groups}\footnote{ This work was partially supported
by CONICET, Argentina}

\bigskip
\bigskip

{\it \large Adri\'{a}n R. Lugo} \\ {\sf
lugo@dartagnan.fisica.unlp.edu.ar}
\bigskip

{\it Departmento de F\'\i sica, Facultad de Ciencias Exactas \\
Universidad Nacional de La Plata\\ C.C. 67, (1900) La Plata,
Argentina}
\bigskip
\bigskip

\end{center}

\begin{abstract}

%{\bf Abstract}
We obtain local parametrizations of classical non-compact Lie groups where
adjoint invariants under maximal compact subgroups are manifest.
Extension to non compact subgroups is straightforward.
As a by-product  parametrizations of the same type are obtained for compact groups.
They are of physical interest in any theory gauge invariant under the adjoint action,
typical examples being the two dimensional gauged Wess-Zumino-Witten-Novikov models
where these coordinatizations become of extreme usefulness to get the background fields
representing the vacuum expectation values of the massless modes of the
associated (super) string theory.

\end{abstract}

%\newpage
\bigskip

\section{Introduction}
\cleqn

It is none to say what group theory meant (and means) for theoretical physics during this
century, in particular the theory of continuous groups or Lie groups ( see e.g. \cite{lie1},
\cite{lie2} and references therein).
The properties of their Lie algebras, easier to hand than the groups themselves, define them
locally and in fact most part of the textbooks are dedicated to them \cite{lie3}.
However depending on the problem at hand, explicit parametrizations of the group manifold
become necessary.
Many of them are widely known, example of them the $SU(2)$ or more generally the Euler angles
of orthogonal groups. Of course that we can always write locally a group element as a product
of one-generator exponentials, or simply as the exponential of an arbitrary Lie algebra element.
But in most cases theses obvious parametrizations are of few usefulness because they obscure
the global properties of the group and generally lead to un-tractable computations.

An interesting parametrization is suggested by the Mackey theorem, the well-known coset
decomposition: let $G$ be a group and $H$ a subgroup of it.
Then for any $g\in G$,
\be
g = k\; h ,  \;\; k\in G/H\; , \;\; h\in H \label{coset}
\ee
in a unique way.
This leads to the theory of homogeneous spaces (or (left or right) coset spaces) $G/H$,
good references on the subject being \cite{hel}, \cite{gil}.
Among others physical applications \eqn{coset} is fundamental in the treatment of
effective field theories with spontaneously broken symmetries \cite{wein}.

But let us assume instead that we have a field theory including maps from ``space-time'' on
a group manifold $G$ among its degrees of freedom, and gauge invariant under the adjoint action
of a subgroup $H$ of $G$
\be
g\;\rightarrow\; {}^h g =  h\; g\; h^{-1}\;\; ,\;\; h\in H  \label{adja}
\ee
This means that effectively the theory depends on the invariants of the group under the adjoint
action of the subgroup.
That is, if we were able to write uniquely any element of $G$ as
\be
g \equiv h^{-1}\; \bar g\; h \;\; , h\in H \label{adj}
\ee
then clearly making a gauge transformation \eqn{adja} identifying the $h$'s the theory will
depend only on $\bar g$ that encloses the invariants  mentioned above.
We would like to remark that at difference of \eqn{coset} there no exists any general theorem
assuring the decomposition \eqn{adj}; in fact it is not difficult to find examples where it
is not possible to write it.

It is the aim of this paper to get the class of local parametrizations of the type \eqn{adj}
in a whole set of cases of physical interest.
Specific examples where they must be used (and were used in the lower dimensionality
cases where parametrizations were available) are the two dimensional gauged
Wess-Zumino-Witten-Novikov models \cite{gwzw}, \cite{gaw}.
It is worth to say however that the results, being explicit parametrizations of
classical groups, are valuable on their own right independently of the applications.

%%%%%%%%%%%%%%%%%%%%%%%%%%%%%%%%%%%%%%%%%%%%%%%%%%%%%%%%%%%%%%%%%%%%%%%%%%%%
\section{The orthogonal groups}
\cleqn

We consider in this section the pseudo-orthogonal groups, $G
\equiv SO(p,q)$. Its maximal compact subgroup is $H\equiv
SO(p)\times SO(q)$. In order to get its decomposition \eqn{adj} we
need as a first step to get the

\subsection{Reduction of $SO(p+1)$ under $SO(p)$.}

We start by writing \cite{gil}
\be
P_{p+1}(\vu_p , P_p ) = K_p(\vu_p )\; H(P_p, 1)\label{cosort}
\ee
where $\;\vu_p\;$ is a $p$-dimensional real vector, $P_p\in SO(p)$ and
generically we will mean
\be
H(P,Q) =  \matriz{P}{0}{0}{Q}
\ee
In what follows the dimensionalities of matrices should be understood from the context
when not stated explicitly.
The right coset element in $SO(p+1)/SO(p)\sim S^p$ is given by
\bea
     K_{p+1}(\vu_p ) &=& \left( \matrix{
               1 - (1 + u_{p+1} )^{-1}\; \vu_p \vu_p{}^t    &\vu_p\cr
               -\vu_p{}^t              &u_{p+1}\cr } \right)\cr
     1 &=& \vu_p{}^t \vu_p + (u_{p+1})^2
\eea
Under an adjoint transformation
\be
{}^h P_{p+1} = H(h,1)\; P_{p+1}\; H(h^t ,1)\;\; ,\;\; h\in SO(p)
\ee
the parameters of $P_{p+1}$ get transformed as:
\bea
{}^h \vu_p &=& h \; \vu_p \cr
{}^h P_p &=& h\; P_p \; h^t
\label{adjtras1}
\eea
The procedure will be constructive.
Les us pick an arbitrary matrix $\;V_p\in SO(p)\;$ decomposed this time as a left
coset w.r.t. $SO(p-1)$
\be
V_p = H(V_{p-1}, 1) \; K_p(\vv_{p-1}) \label{parV}
\ee
and rewrite \eqn{cosort} for any such a $V_p$ with the help of \eqn{adjtras1} as
\bea
P_{p+1} &=& H(V_p{}^t ,1)\; {\bar P}_{p+1} \; H(V_p , 1)\cr
{\bar P}_{p+1} &=& K_{p+1}(V_p\vu_p )\; H(V_p P_p V_p{}^t ,1)\label{adjort}
\eea
The general idea to apply here and in the subsequent cases is to fix the whole matrix $V_p$
($\equiv \vv_{p-1}, V_{p-1}$ ) in terms of variables of $P_{p+1}$ ($\equiv \vu_p , P_p$),
leaving aside only precisely the invariants together with the matrix $V_p$ as parameters of
$P_{p+1}$.
Evidently this procedure is equivalent to make a change of variables
from the non-invariant parameters in $P_{p+1}$ to $V_p$.

Equation \eqn{adjort} suggests to put $\;\vu_p\;$ in some standard form by a specific
choice of (a part of) $V_p$.
In fact it is easy to show that the choice
\footnote{ As usual we will use the notation $({\check e}_i )_j = \delta_{ij},
(E_{ij})_{kl}=\delta_{ik} \delta_{jl}, A_{ij} \equiv E_{ij} - E_{ji}, S_{ij}
\equiv E_{ij} + E_{ji} $.}
\be
\left( \begin{array}{c}
\vv_{p-1} \\ - v_p \end{array}\right) = - \frac{\vu_p}{|\vu_p |}\label{cvuno}
\ee
defines the rotation
\be
K_p ( \vv_{p-1})\; \vu_p = | \vu_p | \; \; {\check e}_p
\ee
Note that we have changed the ($p-1$) parameters from $\vu_p$ indicating its direction
in terms of the ($p-1$) parameters in $\vv_{p-1}$.
With such a choice of $\vv_{p-1}$ ($V_{p-1}$ not fixed yet) we can write
\be
{\bar P}_{p+1} = K_{p+1} (| \vu_p | \; {\check e}_p )\;
H( H(V_{p-1},1)\; P_p\; H(V_{p-1},1)^t , 1)
\ee
where we have redefined
$\; P_p\rightarrow K_p (\vv_{p-1})^t\; P_p \; K_p(\vv_{p-1})$.
But according to \eqn{adjort} we can rewrite it as
\be
{\bar P}_{p+1} =
 K_{p+1} (| \vu_p | \; {\check e}_p )\; H({\bar P}_p , 1)
\ee
Inspection of this formula indicates an iterative process, the next step being to write the
analogous expression for ${\bar P}_p$ and so on; after $p$ steps we get
\be
{\bar P}_{p+1} = \prod_{l=1}^{\overleftarrow p}\; H\left( K_{l+1}
( |\vu_l| {\check e}_l ), 1_{p-l} \right)
\ee
It is convenient to introduce the angular variables
\be
|\vu_l | = \sin \theta_l \;\; , \;\; 0 \leq\theta_l\leq \pi
\ee
and write the $SO(p+1)$ element in the final form
\bea P_{p+1} &=& H( V_p , 1)^t\; {\bar P}_{p+1}\; H( V_p , 1)\cr
{\bar P}_{p+1} &=& \prod_{l=1}^{\overleftarrow p}\;
\exp\left( \theta_l\;A_{l,l+1}\right)\label{adjortfinal}
\eea
which displays explicitly the $p$ invariants $\{ \theta_l , l=1,\cdots,p \}$ under
the adjoint action of $SO(p)$.
Note that the number of parameters trivially matches:
${p\over 2} (p-1) + p = {p\over 2} (p+1)$, as should;
this is the first of our results, to be used in the following.

\subsection{Reduction of $SO(p,q)$ under $SO(p)\times SO(q)$.}

Our starting point is again the right coset parametrization \cite{gil}
\footnote{
An explicit derivation from the definition of pseudo unitary
groups is given in Appendix A of \cite{lu1}.
}
\be
\Lambda_{p,q}(S,P_p,Q_q) = K_{p,q}(S)\; H(P_p,Q_q) \label{parortpq}
\ee
where $P_p (Q_q) \in (SO(p)( SO(q) )$ and the coset element is given by
\bea
K_{p,q}(S) &=& \exp\matriz{0}{N}{N^t}{0} =
\matriz{(1 + S S^{t})^{\frac{1}{2}}}{S}{S^{t}}{(1 + S^{t}S)^{\frac{1}{2}}}\cr
S &=& (N N^t )^{-{1\over2}} \sinh (N N^{t})^{1\over 2} N  \;\; \in \Re^{p\times q}
\label{cospseu}
\eea
Under an adjoint transformation
\be
{}^h\Lambda_{p,q} = H(h_p,h_q )\;\Lambda_{p,q}\; H(h_p,h_q )^t
\ee
with $ h_p (h_q)\in SO(p)(SO(q))$, the parameters of $\Lambda_{p,q}$ transforms as
\bea
{}^h S &=& h_p \; S \; h_q{}^t \cr
{}^hP_p &=& h_p \; P_p \; h_p^t\cr
{}^hQ_q &=& h_q \; Q_q \; h_q^t
\eea
By following the strategy pursued in the past subsection we introduce two matrices $V_p, V_q$
belonging to $SO(p), SO(q)$ respectively and rewrite \eqn{parortpq}
\bea
\Lambda_{p,q} &=& \; H(V_p ,V_q )^t\;{\bar\Lambda}_{p,q} \;H(V_p ,V_q) \cr
{\bar\Lambda}_{p,q} &=& exp\matriz{0}{V_p N V_q{}^t}{ (V_p N V_q{}^t
)^t}{0} \; H(V_p P_p V_p{}^t , V_q Q_q V_q{}^t )
\eea
As in \eqn{parV} we consider left coset parametrizations
\bea
V_p = H(V_{p-1}, 1) \;K_p(\vv_{p-1})\cr
V_q = H(V_{q-1},1)\;K_q(\vv_{q-1})
\eea
and try to totally fix $\;\vv_{p-1}\;$ and $\;\vv_{q-1}\;$ to put $N$ in a standard form.
It turns out that it is possible to choose these vectors in such a way that
\be
V_p \; N\; V_q{}^t \equiv H(V_{p-1} , 1)K_p(\vv_{p-1})\; N
K_q(\vv_{q-1})^t\; H(V_{q-1}{}^t , 1) = \matriz{N_r}{0}{0}{n}\label{Nst}
\ee
where $\;N_r \in \Re^{(p-1)\times (q-1)}\;$ and $\;n\in \Re\;$.
In fact if we write
\be
N = \matriz{N'_r}{\vn_{p-1}}{\vn_{q-1}{}^t} {n'}
\ee
then is straightforward to verify that the gauge fixing condition
$\vn_{p-1}= \vn_{q-1} = \vec 0$, i.e.
\be
K_p(\vv_{p-1})\;N K_q(\vv_{q-1})^t = \matriz{N_r}{0}{0}{n}
\ee
holds if we choose
\bea
\vv_{p-1} &\equiv& (1+ |\vt_{p-1}|^2 )^{-\frac{1}{2}}\; \vt_{p-1}\cr
\vv_{q-1} &\equiv& (1+ |\vt_{q-1}|^2 )^{-\frac{1}{2}}\; \vt_{q-1}
\eea
with $\;\vt_{p-1} , \vt_{q-1}\;$ satisfying
\footnote{
This set of equations can be reduced to a system of two (quadratic) equations
with two unknowns; the important thing for us is that solutions
exist and define the change of variables
\begin{eqnarray*}
(\vn_{p-1} ,\vn_{q-1})\rightarrow (\vv_{p-1} ,\vv_{q-1})
\end{eqnarray*}
}
\bea
\vt_{p-1} &=& ( \vn_{q-1}{}^t\vt_{q-1} -
n)^{-1}\;\left( \vn_{p-1} - N'_r \vt_{q-1}\right)\cr
\vt_{q-1} &=& (\vn_{p-1}{}^t\vt_{p-1} - n)^{-1}\left( \vn_{q-1} - N'_r{}^t\vt_{p-1}\right)
\eea
A final redefinition $\;N_r\rightarrow V_{p-1}{}^t\; N_r\; V_{q-1}\;$ leads to \eqn{Nst}.
Finally (after reparametrizing
$\; P_p \rightarrow K_p(\vv_{p-1})^t\; P_p\; K_p(\vv_{p-1})\; ,
\;  Q_q \rightarrow K_q(\vv_{q-1})^t\; Q_q\; K_q(\vv_{q-1})\;$) we use \eqn{adjortfinal}
to fix $V_{p-1}$ and $V_{q-1}$; the result can be recast in the form
\bea
\Lambda_{p,q} &=& \; H(V_p ,V_q )^t\;{\bar\Lambda}_{p,q} \;H(V_p ,V_q) \cr
{\bar\Lambda}_{p,q} &=& \exp\left( \sum_{i=1}^{p-1}\sum_{j=1}^{q-1} N_{ij} \; S_{i,p+j} + n
S_{p,p+q}\right)\;  \prod_{k=1}^{\overleftarrow {p-1}}\; \exp(\theta_k A_{k,k+1})\;
\prod_{k=1}^{\overleftarrow{q-1}}\;\exp(\bar\theta_k A_{p+k,p+k+1})\cr & &\label{adjpseu}
\eea
Again we verify the matching in the number of parameters
$\; V_p, V_q, N_{ij}, n, \theta_k , \bar\theta_k\;$,
\be
{p\over 2} (p-1)+ {q\over 2} (q-1) + (p-1)(q-1) + 1 + (p-1) +
(q-1) = {1\over 2} (p+q) (p+q -1)
\ee
It is worth to say that the first term in \eqn{adjpseu} can be computed as in \eqn{cospseu}
with $N$ as in the r.h.s. of \eqn{Nst}; however this is a formal expression for which, to
our knowledge, only the ``minkowskian'' cases $p=1$ or $q=1$ admit an explicit form.

%%%%%%%%%%%%%%%%%%%%%%%%%%%%%%%%%%%%%%%%%%%%%%%%%%%%%%%%%%%%%%%%%%%%%%%%%%%%%%
\section{The unitary groups}
\cleqn

The treatment of these groups parallels that made in the case of
the orthogonal ones, with some additional complications due to the
complex character of them. As before we start considering

\subsection{Reduction of $SU(p+1)$ under $U(p)$.}

An element of $SU(p+1)$ can be written as
\be
P_{p+1}(\vu_p , U_p ) = K_{p+1}(\vu_p )\; H(U_p, u_p^* ) =
K_{p+1}(\vu_p )\; H(P_p,1)\; \exp\left( i\;\phi^p T_p\right)\label{cosun}
\ee
where
$\; u_p\equiv\det U_p=\exp(ip\phi_p )\;,\;\vu_p\in\bc^p\; ,\; U_p\in U(p),P_p\in SU(p)\;$,
and we have introduced  a convenient basis
$\;\{ T_k = \sum_{l=1}^k ( E_{ll} - E_{k+1,k+1} ),\; k=1,\cdots,p\}\;$
in the Cartan subalgebra of $su(p+1)$.
The right coset element in \eqn{cosun} belonging to $SU(p+1)/U(p)\sim CP^p$
is given by \cite{gil}
\bea
K_{p+1}(\vu_p ) &=&  \matriz{1 - (1 + u_{2p+1} )^{-1}\; \vu_p\vu_p{}^\dagger}
{\vu_p}{-\vu_p{}^\dagger}{u_{2p+1}} \cr
1 &=& \vu_p{}^\dagger \vu_p + (u_{2p+1})^2
\eea
The adjoint action under $\; H_p \in U(p)\;$ is ($\;h_p \equiv \det H_p$)
\be
{}^H P_{p+1} = H( H_p, h_p^*)\; P_{p+1}\; H(H_p ,h_p^*)^\dagger \;\longleftrightarrow\;
\left\{ \begin{array}{l} {}^V\vu_p = h_p\; H_p \; \vu_p \\
{}^H P_p =  H_p\; P_p \; H_p^\dagger \\\; {}^H \phi^p = \phi^p \end{array}\right.
\label{adjtrpsu}
\ee
As before we pick an arbitrary  $V_p \in U(p)$ left coset parametrized
\be
V_p = H( V_{p-1} , v_{p-1}^*)\; K_p (\vv_{p-1})\; \exp( i\beta_p)\;\;\; ,\;\; V_{p-1}\in U(p-1)
\ee
and write \eqn{cosun} with the help of \eqn{adjtrpsu} as
\bea
P_{p+1} &=& H(V_p , v_p^*)^\dagger\; {\bar P}_{p+1} \; H(V_p,v_p{}^*)\cr
{\bar P}_{p+1} &=& K_{p+1}(v_p V_p\vu_p )\; H(V_p P_p V_p{}^\dagger
,1)\; \exp(i\;\phi^p T_p ) \label{adjuni}
\eea
By choosing $\vv_{p-1}$ and ${\beta_p}$ ($V_{p-1}$ free) as
\bea
\left( \begin{array}{c} \vv_{p-1} \\ - v_{2p-1} \end{array}\right) &=&
- \frac{(\vu_p^*)^p}{|(\vu_p)^p |}\; \frac{\vu_p}{|\vu_p |} \cr
\exp (i\beta_p ) &=&
\left( \frac{(\vu_p^* )^p}{|(\vu_p{} )^p |}
\right)^\frac{1}{p+1}\label{cvdos}
\eea
we have
\be
v_p \; V_p \; \vu_p = |\vu_p| \; {\check e}_p
\ee
and identifying $V_{p-1}$ as the $SU(p-1)$ matrix corresponding to the $P_p$ decomposition
in \eqn{adjuni} ( previous redefinition
$\; P_p\rightarrow K_p\; (\vv_{p-1})^\dagger P_p \; K_p(\vv_{p-1})\;$ ) we get
\be
{\bar P}_{p+1} = K_{p+1}(|\vu_p | {\check e}_p )\; H({\bar P}_p ,1)\;\exp(i\;\phi^p T_p )
\ee
By repeating the analysis for ${\bar P}_p $ and after $p$ steps we arrive to the final result
\bea
P_{p+1} &=& H( V_p , v_p^*)^\dagger\; {\bar P}_{p+1}\; H( V_p ,v_p^* )\cr
{\bar P}_{p+1} &=& \prod_{l=1}^{\overleftarrow p}\; \exp\left( \theta_l\;
A_{l,l+1}\right)\; C_{p+1}(\Phi ) \label{adjunifinal}
\eea
It differs from \eqn{adjortfinal} from the unitary character of $V_p$ and the
comparison of the arbitrary Cartan element
$\;C_{p+1}(\Phi )=\exp(i \sum_{l=1}^p \;\phi^l T_l )\;$ at right in ${\bar P}_{p+1}$.

\subsection{Reduction of $SU(p,q)$ under $S(U(p)\times U(q))$.}

In order not to be repetitive we will skip some steps in what follows.
An arbitrary element in $SU(p,q)$ can be left coset decomposed under
$\;S(U(p)\times U(q))\;$ as
\bea
\Lambda_{p,q}(S,U_p,U_q) &=& K_{p,q}(S)\; H(U_p,U_q)\;\;\; ,\;\; u_p = u_q{}^* \cr
K_{p,q}(S) &=& \matriz{(1 + S S^\dagger)^{\frac{1}{2}}}{S}{S^\dagger}{(1 + S^\dagger S)^{
\frac{1}{2}} } = \exp  \matriz{0}{N}{N^\dagger}{0}\cr
S &=& (N N^\dagger )^{-{1\over2}} \sinh (N N^\dagger)^{1\over 2} N \;\;\; ,\;\;
S, N\in {\cal C}^{p\times q}\label{parunipq}
\eea
The adjoint action under $\; H(h_p,h_q ) \in S(U(p)\times U(q) )\;$ is
\be
{}^H \Lambda_{p,q} =  H(h_p,h_q )\;\Lambda_{p,q} H(h_p ,h_q )^\dagger
\longleftrightarrow \left\{\begin{array}{l} \;{}^H S = h_p \; S \; h_q{}^\dagger \cr
{}^H U_p = h_p \; U_p \; h_p^\dagger\cr
{}^H U_q = h_q \; U_q \; h_q{}^\dagger \end{array}\right.
\ee
Two matrices $V_p, V_q$ belonging to $U(p), U(q)$ respectively
with $v_p v_q =1$ are introduced and following similar steps as in
Subsection $2.2$ we find that it is possible to fix $N$ in the way
\be
N = \matriz{N_r}{0}{0}{n}
\ee
where now $N_r \in {\cal C}^{(p-1)\times(q-1)}$ and $n\in \Re $.
Then by using the results of the past subsection we get the final result for the
parametrization
\bea
\Lambda_{p,q} &=& \; H(V_p, V_q)^\dagger \;{\bar \Lambda}_{p,q} \;H(V_p ,V_q) \cr
{\bar \Lambda}_{p,q} &=& \exp \matriz{0}{N}{N^\dagger}{0}\;
\overleftarrow{\prod_{l=1}^{p-1}}\;\exp\left(\theta_l\; A_{l,l+1}\right)\;
\overleftarrow{\prod_{l=1}^{q-1}}\;\exp\left({\bar\theta}_l\; A_{p+l,p+l+1}\right)\;C(\Phi)
%\exp\left( \sum_{i=1}^{p-1}\\ \phi^i T_i ) + \sum_{i=1}^{q-1}{\tilde\phi}^i T_{p+i}\right)\;
%\exp\left( i\; \alpha \matriz{q1_p }{0}{0}{-p 1_q} \right)
\label{adjpsunifin}
\eea
where $\; H(V_p, V_q)\in S(U(p)\times U(q))\;$ and we denote by $\;C(\Phi)\;$ an arbitrary
element in the Cartan subalgebra of the Lie algebra of $S(U(p)\times U(q))$.

\subsection{Decomposition under the maximal torus}

Some times is useful to have the adjoint decomposition of unitary groups under the
Cartan subalgebra.
We will work out for definiteness the case of $SU(n+1)$; the non compact versions differ
as usual by signs in the coset elements and Wick rotations of some compact generators.

To this end we begin by searching for the coset decomposition of $SU(n+1)$ under
$C(SU(n+1))$; from \eqn{cosun}
\bea
P_{n+1}(\vu_n , U_n ) &=& K_{n+1}(\vu_n )\; H(U_n, u_n^* )\cr
      K_{n+1}(\vu_n ) &=&  \exp\theta_n\matriz{0}{\check{r}_n}{-\check{r}_n{}^\dagger}{0}
= \matriz{1 - (1 + u_{2n+1} )^{-1}\;
\vu_n\vu_n{}^\dagger}{\vu_n}{-\vu_n{}^\dagger}{u_{2n+1}}\cr
|\vu_n|^2 + (u_{2n+1})^2 &=& 1 \;\;\;, \;\;\vu_n = \sin\theta_n \;{\check r}_n\;\;\; ,
\;\;{\check r}_n{}^\dagger {\check r}_n =1\;\;\; ,\;\; \theta_n \in [0,\pi]\label{ec}
\eea
By introducing
\bea
U_n &=& P_n\;\matriz{1_{n-1}}{0}{0}{u_n} \;\; , \;\;  P_n \in SU(n)\cr
u_n &\equiv& \det U_n = \exp(i\varphi^n)
\eea
we can rewrite \eqn{ec} as
\be
P_{n+1}(\vu_n ,U_n ) = K_{n+1}(\vu_n )\; H(P_n ,1)\;\exp(i\;\varphi^n H_n)
\ee
where this time is convenient to introduce the basis
$\;\{ H_k = E_{kk} - E_{k+1,k+1} ,k=1,\cdots, n\}\;$ in the Cartan subalgebra of
$su(n+1)$.
By repeating with $P_n$ and iterating we get
\be
P_{n+1} = \overleftarrow{\prod_{l=1}^n} \;
\matriz{K_{l+1}(\vu_l)}{0}{0}{1_{n-l}} \;\exp(i\;\vec\varphi \cdot {\vec H})
\ee
Now let us pick an element $C_{n+1}(\vec\alpha )\equiv \exp (\alpha^i H_i)\in
C(SU(n+1))$ and write as usual
\bea
P_{n+1} &=& C_{n+1}(\vec\alpha )^\dagger \;\overleftarrow{\prod_{l=1}^{n}}
\left( C_{n+1}(\vec\alpha )\;\matriz{K_{l+1}(\vu_l)}{0}{0}{1_{n-l}}\;
C_{n+1}(\vec\alpha )^\dagger\right)\; C_{n+1}(\vec\varphi)\;C_{n+1}(\vec\alpha )\cr
& &
\label{unicarini}
\eea
We see that the $(\varphi^l )$ variables are invariant; we must choose
the ${\alpha^i}$'s to kill parameters in the productory.
It is easy to show that the versors $\check{r}_l$ get transformed
as
\be
{}^\alpha\check{r}_l = \exp (-i\tilde\alpha_{l+1} )\; \exp(
i\sum_{i=1}^{l}\tilde\alpha_i E_{ii})\;\check{r}_n\;\;\; ,\;\; l=1,\cdots,n
\ee
where
\be
{\tilde\alpha}_k =
\left\{ \begin{array}{ll}
\alpha_1 &  \;\; if\;\; k=1\cr
 -\alpha_{k-1} + \alpha_k & \;\; if\;\;  k=2,\ldots,n \cr
-\alpha_n & \;\; if\;\; k=n+1
\end{array} \right.
\ee
Then we can put the phases of the $({}^\alpha\check{r}_l )^l$ components, $l=1,\cdots,n$,
to zero by choosing the $\alpha's$ such that
\footnote{Equations \eqn{kcm} can be formally solved by $\;\;  \alpha^i =
- {K^{-1}}^{i}{}_j \; phase(\check{r}_j )^j$ where $K$ is the
Killing-Cartan matrix of the $A_n$ algebra; it is probable that this fact
does not be an accident but occurs in other cases $G/C(G)$.
}
\bea
2 \alpha_1 -\alpha_2 &=& - phase(\check{r}_1 )^1\cr
- \alpha_1 + 2\alpha_2  -\alpha_3 &=& -phase(\check{r}_2)^2\cr
\vdots & &\vdots\cr
 -\alpha_{n-1}+2\alpha_n &=& -phase(\check{r}_n )^n \label{kcm}
\eea
In other words, from \eqn{unicarini} we get the final result
\bea
P_{n+1} &=& C_{n+1}(\vec\alpha)^\dagger\; {\bar P}_{n+1}\; C_{n+1}(\vec\alpha)\cr
{\bar P}_{n+1} &=& \overleftarrow{\prod_{l=1}^{n}}
\matriz{K_{l+1}(\vu_l)}{0}{0}{1_{n-l}}\;C_{n+1}(\vec\varphi )\label{unicarfin}
\eea
with the constraints implied by \eqn{kcm}: $\;(\check{r}_l)^l \in \Re \; ,\;
l=1,\ldots, n$.

\section{The decomposition of $Sl(n)$ under $SU(n)$.}
\cleqn

This is one of the two irreducible riemannian cases
\footnote{
The other one is $SU^*(2n)/USp(2n)$ and will not be considered here.}
in the sense that the coset element is not of the form
$\exp\matriz{0}{N}{\pm N^\dagger}{0}\;$ for some matrix $N$ (off-diagonal cosets).
We start from the well-known coset decomposition under $SO(n)$ of
any unimodular real $n\times n$ matrix
\be
g_n = S_n \; P_n\;\; ,\; S_n{}^t = S_n \;\;\; ,\;\; P_n \in SO(n)
\ee
Also $S_n$ is positive definite and $\det S_n = 1$.
But we know from elementary linear algebra that any such a matrix is
diagonalizable by an orthogonal one $Q_n$ completely determined
\bea
S_n &=& Q_n{}^t \; Diag(\lambda_1{}^2 , \ldots, \lambda_n{}^2 ) \; Q_n\cr
\prod_{i=1}^{n}\lambda_i &=& 1\;\;\; ,\;\;
\lambda_1\geq \lambda_2\geq \dots \geq\lambda_n\geq 0
\eea
from where after a redefinition $P_n \rightarrow  Q_n{}^t P_n Q_n$ we get
\bea
g_n &=& Q_n{}^t \;{\bar g}_n \; Q_n \;\;\; , Q_n\in SO(n)\cr
{\bar g}_n &=& Diag(\lambda_1{}^2 , \ldots, \lambda_n{}^2 )\;
P_n \label{slnort}
\eea
Analogous steps using well known results yield the complexification of \eqn{slnort},
namely the decomposition of $Sl(n, \cal C)$ under $SU(n,\cal C)\;$ that we
quote without proof
\bea
g_n &=& V_n{}^\dagger \;{\bar g}_n \; V_n \;\;\; , \;\; V_n \in SU(n, \cal C )\cr
{\bar g}_n &=& C_{n}(\vec\alpha) \; \overleftarrow{\prod_{l=1}^{n-1}}
\matriz{K_{l+1}(\vu_l )}{0}{0}{1_{n-1-l}}\; C_{n}(\vec\beta)\label{slnuni}
\eea
where $\; C_n \in C(SU(n,{\cal C} ))\;$ and the vectors $\vu_l$ are constrained by
$\; (\check{r}_l)^l \in \Re\; $ as in Section 3.3.

\section{Conclusions}

We have obtained in this paper adjoint parametrizations defined by \eqn{adj} w.r.t.
maximally compact subgroups for a large set of non-compact groups, the classically
riemannian cosets.
The constructive procedure used allows to extend them straightforwardly to non riemannian
decompositions, for example $\;SO(p+n,q)\;$ under $\;SO(n,q)\;$.
Few words about symplectic groups: these groups can be treated in the same way as made here;
the fact that $\; USp(2p, 2q) \sim U(2p, 2q; {\cal C} )\bigcap Sp(2p+2q;{\cal C})\;$
seems to suggest that the corresponding decomposition under $\;USp(2p)\times USp(2q)\;$
could follow from replacing in \eqn{adjpsunifin} $E$  by $Z$ generators
\footnote{
See reference \cite{gil}, chapter 5 for definitions.
}
and respective Cartan subalgebras, but we have not checked this.

We remark the locality of the parametrizations obtained; the changes of variables needed
to carry out the job are singular in some points of the group manifold as can be seen by
direct inspection of \eqn{cvuno} , \eqn{cvdos} for example.

Possible applications in physical problems of these parametrizations are in the context of
GWZW models as models of strings moving on background fields.
Equation \eqn{adjpseu} can be used to treat systematically all the models in \cite{bs1}
where the lowest dimensional cases were considered.
Also the measure on the group can be computed straightforwardly through the Maurer-Cartan
forms without introducing Fadeed-Popov ghost due to constraints because they were solved
once for all.
For $p+q= 2$ ($A_1$ algebras) parametrization \eqn{adjpsunifin} is widely known; for
$p=2, q=1$ was introduced in \cite{lu1}; it allows to extend the study of coset models in the
search of physically relevant string backgrounds represented by exact conformal field
theories.

\section*{Acknowledgements} It is a pleasure to thank to Loriano Bonora for useful
correspondence.
%%%%%%%%%%%%%%%%%%%%%%%%%%%%%%%%%%%%%%%%%%%%%%%%%%%%%%%%%%%%%%%%%%%%%%%%%%

\end{document}